# Economic and Environmental Sustainability Through Reshoring: A Case Study

Shaikh, MD Parvez, and Sarder, MD

*Abstract*— Not too long ago, offshoring was considered a panacea for many U.S. Companies to achieve economic sustainability. Offshoring also created an unnecessary movement of goods between the point of consumption and the point of sourcing and hence contributed to greenhouse gas emissions. With many things changed, hundreds of U.S. companies have started Reshoring. Due to supply chain disruptions and increased tax implications, including tariffs, there is a growing desire among companies to achieve economic and environmental sustainability through reshoring. This model case study highlighted the common offshoring challenges and demonstrated new methods/solutions for the companies to save their bottom line. Using the Reshorability Index (RI) and Total Cost of Ownership (TCO) we developed a model to show which products or components we should bring back to the U.S. instead of continuing offshoring. From this study, we have found out that reshoring is not only an economically profitable decision but also has a positive impact on reducing GHG (Greenhouse Gas) emissions. Our research found that the companies that currently offshore heavy products will benefit more from implementing our developed model. Leveraging this model, industries can identify, and compare the ownership cost of their purchased materials and take the decision on potential reshoring. Additionally, companies will be able to calculate the GHG emission and identify the reduction of such emissions due to reshoring.

*Note to Practitioner*— This research developed ready-to-use models for any practitioner or decision maker to assess whether a product or component should be reshored or not based on the reshorability index and total cost of ownership. These models are verified and validated by a real U.S. manufacturer. Reshorability Index (RI) is a quantitative measure of a product's favorability to be reshored. For instance, if the RI of a product is 75%, the company will most likely enjoy significant cost benefits from outsourcing that product from a local manufacturer. The total cost of ownership (TCO) model was originally developed by Harry Moser and used in this research. The total cost of ownership (TCO) is an analysis that looks at the hidden costs beyond FOB price. The TCO includes various costs associated with offshoring such as logistics cost, packaging cost, and intellectual property cost. Both models are easy to implement, and they provide critical information about the reshoring decision. The other models used in this paper are very helpful to quantify the greenhouse gas reduction related to reshoring decisions.

## I. INTRODUCTION

In previous decades, the U.S., as well as other industrialized countries' companies, transferred (a.k.a. offshored) manufacturing operations (or other business processes) to low labor-cost countries, mainly in Asia. This offshoring and outsourcing trend (handing a business process to an external service provider) towards low-cost manufacturing destinations, combined with enhanced ocean shipping and improved onshore and inland intermodal services, constituted one of the most significant changes in manufacturing and supply chain strategy around the world. Offshoring gradually transformed the global manufacturing environment, in which fixation on low-cost labor was and most probably still is, a dominant motive for the manufacturing location decision. This offshoring phenomenon was followed by a decline in U.S. manufacturing and its share in the nation's GDP, but also by millions of lost jobs this sector, traditionally, offered.

However, after nearly a quarter of a century, the offshoring trend, that reduced the manufacturing sector in most developed countries, seems to have reached an inflection point. Publicity on the phenomenon, but also the realistic need for a better understanding of sourcing and global manufacturing trends in general, led many researchers in the U.S. and around the world, to further study and investigate the reshoring manufacturing phenomenon, its potential, and applicability. Furthermore, to better capture the reshoring phenomenon and quantify potential reshoring benefits, various researchers tried to provide decision-makers with more elegant evaluation tools, and cost estimation models as well as to give a definitional background and framework analysis of the location decision.

Surveys revealed that U.S.-based companies have already started to take under consideration new sourcing destinations, as offshoring production to traditional low-cost countries like China, is not as profitable as it used to be. At the same time, other sourcing options, like Bangladesh, are already experiencing wage increases, while lacking infrastructure production quality, but most of all, unable to reach the necessary production scales to be viable substitutes. Besides production costs, other explicit or implicit costs and risks of transportation and supply chain costs (e.g., disruption, complexity related, etc.), hidden costs (e.g., political uncertainty, corruption, intellectual property theft, etc.) and various other costs, that will be thoroughly examined within the context of this review, but also a trend that urges supply to be as close as possible to the demand, has led U.S. based companies to consider relocating production through reshoring, nearshoring (relocating manufacturing to a country that is close to the U.S.) or onshoring (relocating manufacturing operations, that would otherwise be offshored, to U.S. States in which production cost is lower than that of the State in which the company is based) Surveys reveal that apart from the drivers that relate to the inability of the offshore location to remain a competitive option (like those previously mentioned), reshoring drives exist that relate to the increasing appeal of U.S. itself as a manufacturing location. Improved speed to market, reduced energy cost, political stability, lower freight costs, less complex and easier to manage supply chains, improved customer service, etc., were the most attractive advantages expected from reshoring/nearshoring. However, the vast majority of relevant research suggests that there are several challenges that all stakeholders and policymakers, will most probably come across and that will have to consider, to provide solid ground and prospective for reshoring manufacturing in the future. Such challenges are the lack of skilled personnel and the need for educational reforms that will provide a skilled workforce to the manufacturing sector; a more attractive taxation and regulatory legislation environment and other incentives that can attract offshore production back to the U.S.

Within the context of this study, an effort has been made to provide definitional and framework-establishing studies, as well as modeling efforts of the costs involved with the location decision, aiming to support decision makers with a consistent methodology and background on their relative, current, or future, projects. Accordingly, reports that various consulting organizations have conducted and articles by industry professionals are presented to give a possible, up-to-date, and in-depth perspective on the reshoring phenomenon, urging all interested counterparts in the reshoring topic, to further investigate sources, as the reshoring phenomenon is ongoing and increasingly attracts the attention of the industrial and research community. Following is a highlight of the main insights.

- Reshoring trends at a very early and evolutionary state to produce safe assumptions [1]. Reshoring must convince that "it is more than a public relations stunt.
- Companies considering outsourcing production systems to countries with low labor costs strongly hold to examine decisive effects like variable manufacturing costs thoroughly [2].
- Manufacturing is not migrating from China as rapidly as most experts expect, despite rising wages at nearly 20% a year [3].
- It will take more than just equalizing labor costs with China and other developing countries to lure manufacturers back to the U.S. [4].
- Wages and productivity are the two most significant factors that are causing many American, and even some foreign companies, to consider moving their manufacturing operations to the U.S. [5].
- The average wage in China increased by 150% from 1999 through 2006 [6].
- The manufacturing cost advantage of China over the U.S. is shrinking fast [7].
- Chinese currency appreciated 30% in the last decade relative to the U.S. currency [8].
- U.S. Government should reform taxes and other costs (e.g., tariffs, duties, healthcare) and minimize regulatory burdens to increase the attractiveness of reshoring [1].
- Even if all landed and hidden costs are identified, there are still business risks that must be considered when offshoring production [9].
- Companies expecting reshoring of production facilities back to the U.S. must consider the following seven factors: Transportation and Energy Costs, Exchange Rates, U.S. Demand, U.S. Talent, Availability of Capital, Tax and Regulatory Climate, and U.S. Labor Costs [10].
- The steady changes in wages, productivity energy costs, currency values, and other factors in Latin America, Eastern Europe, and most of Asia are redrawing the map of global manufacturing cost competitiveness [11]. U.S. and Mexico follow, in terms of attractiveness, the pattern of "Rising Global Stars"
- Total Cost of Ownership (TCO)[1] Estimator helps firms to calculate all costs associated with sourcing parts, components, and finished products from different locations. According to the Reshoring Initiative data, 1,312,000 jobs have been brought back to the USA since the manufacturing employment law of 2010 through December 202.

A reshoring strategy can help companies to minimize the high tariffs and logistics costs, improve their supply chain performance and reduce the overall component's cost. Reshoring refers to the practice of transferring a business operation that was moved overseas back to the country from which it was originally relocated. It is nothing but bringing operations back to the United States. In the past few decades, numerous companies in the United States as well as other industrialized countries offshored manufacturing operations (or other business processes) to low-labor-cost countries, mainly in Asia. This offshoring and outsourcing trend (i.e., handing a business process to an external service provider), which is based on low-cost manufacturing destinations, combined with enhanced ocean shipping and improved onshore and inland intermodal services, would constitute one of the most significant changes in manufacturing and supply chain strategy around the world. There are numerous reasons for companies to reshore including shorter lead time, better quality control, rising wages in developing countries, better protection of intellectual property, lower energy cost, and lower freight costs. (Logistics Transportation system)

Some factors may discourage offshoring and motivate a

---

[1] The Total Cost of Ownership (TCO) Estimator is a free online tool that helps companies account for all relevant factors — overhead, balance sheet, risks, corporate strategy and other external and internal business considerations — to determine the true total cost of ownership. https://reshorenow.org/tco-estimator/

company to reshore processes that it had previously outsourced. The reasons that Boston Consulting Group[2] gave were labor cost, product quality, ease of doing business, and proximity to customers. The following are some top issues that may drive companies to reshore their products in the US. Customer retention, Ease of doing business, Policy regulations, Labor cost, Logistics cost, Product Quality, Proximity to customers, Tax implication (including tariff), Incentive/subsidy, and Better Supply Chain Management

Harry Moser[3] suggested that fewer imports and more exports should occur to reduce the U.S. trade deficit sufficiently to increase the nation's GDP by several percentage points. He suggested that instead of depending on a strategy of increasing exports, increasing domestic production by reshoring the manufacturing of products consumed in the U.S. would displace imports and increase the range of goods to be exported. According to the author, over the past decade, about six million manufacturing jobs were lost [12], partially due to offshoring. Offshoring was the result of multiple factors that led to a decline in U.S. competitiveness. Some of the factors, like low wages and lack of regulation in developing countries, could not be controlled that recruited foreign investment and provided companies with strategic growth markets. Moreover, email and teleconferencing enabled efficient communication between the U.S. and factory managers abroad. The effectiveness of sending product designs and manufacturing-process specifications to foreign factories via the Internet, together with the upsurge of cheap container shipping, strengthened the case for offshoring. The author also mentioned that the main error made by companies when they offshore production was focusing solely on low labor rates and product prices. He points out that poor recruitment and training of a skilled manufacturing workforce has become the most important controllable factor influencing U.S. offshoring and reshoring trends. Moreover, the U.S. is the target market for exports from all foreign countries and companies. The author mentions that U.S. companies are still trying to understand the hidden costs related to offshoring. He mentions a new tool called the Total Cost of Ownership (TCO) Estimator, devised by the Reshoring Initiative, which helps firms to calculate all costs associated with sourcing parts, components, and finished products from different locations. He approves that when companies understand the true costs of offshoring, they shall offshore less and reshore more.

After using the Reshoring Initiative's TCO analysis program on data from various companies, it is found that about 25% of the production that has departed overseas might come back to the U.S. if companies consider factors other than the purchase price or landed cost while making sourcing decisions. Moreover, many published surveys also confirm the reshoring trend. Companies are increasingly recognizing that costs, risks, and strategic impacts previously ignored are large enough to overcome the shrinking emerging market wage advantages. They are seeing the benefits of proximity, i.e., producing in the market, especially when the home market is the U.S., still the world's largest.

Another possibility arises here to compare the greenhouse gas emission from the previous offshore supply chain to the newly developed one. The Greenhouse Gas Protocol Initiative[4] provides a detailed method to study the emission including scopes: 1, 2, and 3. It launched in 1998, and the Initiative's mission is to develop internationally accepted greenhouse gas (GHG) accounting and reporting standards for businesses and to promote their broad adoption. This Initiative is a multi-stakeholder partnership of businesses, non-governmental organizations (NGOs), governments, and others convened by the World Resources Institute (WRI), a U.S.-based environmental NGO, and the World Business Council for Sustainable Development (WBCSD), a Geneva-based coalition of 170 international companies. The GHG Protocol Initiative comprises two separate but linked standards [13]: (1) GHG Protocol Corporate Accounting and Reporting Standard. (2) GHG Protocol Project Quantification Standard. This GHG Protocol Corporate Standard provides standards and guidance for companies and other types of organizations preparing a GHG emissions inventory. It covers the accounting and reporting of the six greenhouse gases covered by the Kyoto Protocol—carbon dioxide ($CO_2$), methane ($CH_4$), nitrous oxide ($N_2O$), hydrofluorocarbons (HFCs), perfluorocarbons (PFCs), and sulfur hexafluoride ($SF_6$). The standard and guidance were designed with the following objectives in mind [14]: (1) To help companies prepare a GHG inventory that represents a true and fair account of their emissions, through the use of standardized approaches and principles. (2) To simplify and reduce the costs of compiling a GHG inventory. (3) To provide businesses with information that can be used to build. (4) An effective strategy to manage and reduce GHG emissions (5) To provide information that facilitates participation in voluntary and mandatory GHG programs. (6) To increase consistency and transparency in GHG accounting and reporting among various companies and GHG programs.

Both businesses and other stakeholders benefit from converging on a common standard. For businesses, it reduces costs if their GHG inventory is capable of meeting different internal and external information requirements. For others, it improves the consistency, transparency, and understandability of reported information, making it easier to track and compare progress over time. This standard is written primarily from the perspective of a business developing a GHG inventory [15]. However, it applies equally to other types of organizations with operations that give rise to GHG emissions, e.g., NGOs, government agencies, and universities. Policymakers and architects of GHG programs can also use relevant parts of this standard as a basis for their accounting and reporting requirements.

For effective and innovative GHG management, setting operational boundaries that are comprehensive concerning

---

direct and indirect emissions will help a company better manage the full spectrum of GHG risks and opportunities that exist along its value chain. Direct GHG emissions are emissions from sources that are owned or controlled by the company. Indirect GHG emissions are emissions that are a consequence of the activities of the company but occur at sources owned or controlled by another company.

To help delineate direct and indirect emission sources, improve transparency, and provide utility for different types of organizations and different types of climate policies and business goals, three "scopes" (scope 1, scope 2, and scope 3) are defined for GHG accounting and reporting purposes. Scopes 1 and 2 are carefully defined in this standard to ensure that two or more companies will not account for emissions in the same scope. Direct GHG emissions occur from sources that are owned or controlled by the company, for example, emissions from combustion in owned or controlled boilers, furnaces, vehicles, etc.; emissions from chemical production in owned or controlled process equipment. Direct $CO_2$ emissions from the combustion of biomass shall not be included in scope 1 but reported separately. GHG emissions not covered by the Kyoto Protocol [5], e.g. CFCs, NOx, etc. shall not be included in scope 1 but may be reported separately [14].

Scope 2 accounts for GHG emissions from the generation of purchased electricity consumed by the company. Purchased electricity is defined as electricity that is purchased or otherwise brought into the organizational boundary of the company. Scope 2 emissions physically occur at the facility where electricity is generated.

Scope 3 is an optional reporting category that allows for the treatment of all other indirect emissions. Scope 3 emissions are a consequence of the activities of the company, but occur from sources not owned or controlled by the company. Some examples of scope 3 activities are extraction and production of purchased materials; transportation of purchased fuels; and use of sold products and services. Although scope 3 is optional, it provides an opportunity to be innovative in GHG management. Companies may want to focus on accounting for and reporting those activities that are relevant to their business and goals, and for which they have reliable information.

This section provides an indicative list of scope 3 categories and includes case studies on some of the categories. Some of these activities will be included under scope 1 if the pertinent emission sources are owned or controlled by the company (e.g., if the transportation of products is done in vehicles owned or controlled by the company). To determine if an activity falls within scope 1 or scope 3, the company should refer to the selected consolidation approach (equity or control) used in setting its organizational boundaries: (1) Extraction and production of purchased materials and fuels. (2) Transport-related activities. (3) Transportation of purchased materials or goods, (4) Transportation of purchased fuels. (5) Employee business travel. (6) Employees commuting to and from work.

(7) Transportation of sold products. (8) Transportation of waste

## II. PROBLEM FORMULATION

The whole model consists of three major steps. First, the reshorability index is formulated and identified based on the industry NAICS code[6]. Three factors contribute to the reshoring decision the RI, export-import deficits, and the associated logistics costs, bringing those items back to the US soil. In the second step, this research identifies the total cost of ownership (TCO) of a particular item or product group. If the second step shows a positive result in terms of economical means. The third step finds out if it is feasible to reshore those items back by minimizing the greenhouse gas emission.

### A. Reshorability Index (RI)

Reshorability Index describes the potential of bringing manufacturing back to US soil. In the research of Sarder [16], they identified 44 subfactors, which are influencing 13 location factors related to eight reshoring factors. There are socioeconomic factors (subfactors) that influence the location decision for manufacturing. These factors are related to the reshoring factors that drive the reshoring decision.

TABLE I
RESHORING FACTORS WITH THEIR SUBFACTORS (ONLY)

| SL. No. | Reshoring factors | Factors Influence Location Factors (Subfactors) |
|---|---|---|
| 1 | Labor Cost, Availability & Skill | Labor Cost |
| | | Availability of skilled labor and talent |
| 2 | Availability of Natural Resources | Access to natural resources |
| 3 | Incentives | Incentives, Tax savings |
| 4 | Policy Regulation/IP Right | Government effectiveness |
| 5 | Proximity to Customers | Size of the local market |
| | | Access to the International and Local market |
| | | Growth of market |
| 6 | Infrastructure | Infrastructure |
| 7 | Ease of Doing Business | Follow your competitor |
| | | Stable and business-friendly environment |
| | | Access to capital market |
| 8 | Presence of Suppliers and Partners | Supply chain efficiency and resiliency |

At this stage of this study, we would like to know which component groups are feasible for reshoring based on the RI.

---

5 The Kyoto Protocol is an international treaty aimed at addressing global climate change by reducing greenhouse gas emissions. It was adopted in Kyoto, Japan, in December 1997, and entered into force in February 2005.

6 NAICS stands for North American Industry Classification System. It is a system used by government agencies in Canada, Mexico, and the United States to classify business establishments into industries based on their primary economic activity.

Earlier in this study, we identified the component with tariff issues falling under some groups representing different industries. For example, Aluminum Casting, Metal Stamping, and all those different groups representing different NAICS codes. The trade deficit (the gap in export and import), and CIF (Cost, Insurance, Freight) cost also has importance in deciding to find a suitable candidate for reshoring. According to this research, though the export and import value have no direct impact on Reshorability Index, these are important for analysis. The industries with a low Reshorability Index are likely to have a high trade deficit (high import but low export). At the same time the industries that have high value ($) of import, will have a higher impact on the economy if brought back to the United States. This research established four steps to calculate the Reshorability Index: Step 1: Selecting socioeconomic factors, Step 2: Reshoring factors, Step 3: Weighting the factors, Step 4: Calculating the Reshorability index

CIF cost has a big impact on the Reshorability Index. If the logistics cost is high, the Reshorability Index is likely to be high also. But there are some exceptions as well. The industries with a higher Reshorability Index have less trade deficit than industries with a comparatively lower Reshorability Index. Calculating Reshorabilty Index and Required Formula:

Step 1: Selecting socioeconomic factors
These are well-accepted indicators of the country's socioeconomic status published by the United Nations, World Bank, World Economic Forum, KPMG, US Census, US Department of Commerce, Boston Consultancy Group Economic Intelligence Unit, etc. All the indicators were Normalized under a 1-7 scale with the Mini-Max formula.

$$Normalized\ score = 6 \times \frac{Country\ score - Minimum\ score}{Maximum\ score - Minimum\ score} + 1 \quad (1)$$

Step 2: Reshoring factors
The importance of these factors is taken from the United Nations Conference on Trade and Development (UNTCAD[7] 2009-11). Where it explains how different factors play a role in selecting the location for different industries. For instance, skilled labor is more important in electrical equipment industries than in chemical industries.

Step 3: Weighting the factors

$$The\ score\ for\ the\ US = \frac{(\sum_{j=1}^{j=m}((\sum_{i=1}^{i=n}S_i)/n))_j \times W_j)}{m} \quad (2)$$

$$The\ score\ for\ China\ or\ other\ Asian\ countries =$$
$$= \frac{(\sum_{j=1}^{j=m}((\sum_{i=1}^{i=n}S_i)/n))_j \times W_j)}{(1-(L_c + mc_L))} \quad (3)$$

where $S_i$ = subfactor from step 1; n = number of subfactors impacting the location factors; W = weight of the location factor for a particular industry from step 2; m = number of location factors; $L_c$ = customs, insurance, Freight cost (%) paid in 2014, for importing based on NAICS code data from the US census; $C_L$ = cost of import duties and inventory for long lead-time from China, considered as 3%.

Step 4: Reshorability index
After applying (2) for logistics cost below formula is applied to develop Reshorability Index.

$$Reshorability\ index\ (RI) =$$
$$\frac{US\ score\ from\ Eq\ (2) - Asian\ country\ score\ from\ Eq\ (3)}{Asian\ country\ score\ from\ Eq\ (3)} \times 100 \quad (4)$$

B. Total Cost Of Ownership (TCO)

Generally, companies are prone to offshore, they make their decision based on the purchase price of the product, the global supply chain is also made an easy choice for them, as all the parts of the world are connected through strong infrastructure. However, making the decision solely on the purchase price is misleading [17]. Harry Moser finds out thirty-five other cost and risk factors beyond the purchase price, if those cost components were taken into consideration the purchase decision might have been changed. Here in this total cost of ownership calculation, he pulled all the cost factors into six different cost buckets: (1) Purchase Price, (2) Cost of Goods Sold (COGS), (3) Other Hard Cost Factors, (4) Risk Costs, (5) Strategic Costs, and (6) Green (Environmental/Sustainability) Cost Factors.

The top reasons that companies have reshored since 2010 include: (1) Lead time, (2) Higher product quality and consistency, (3) Rising offshore wages, (4) Skilled workforce, (5) Local tax incentives, (6) Image of being Made in the USA, (7) Lower inventory levels, better turns, (8) Better responsiveness to changing customer demands (9) Minimal intellectual property and regulatory compliance risks (10) Improved innovation and product differentiation

In the last 2 to 3 years the costs of extended supply chains have risen dramatically, especially sea and air freight and Section 301 tariffs [18], [19]. Meanwhile, the risks of supply chain disruption have increased even faster. The trade war, COVID, and Suez Canal blockage [20] have shown what can happen. Now the Ukraine/Russian war [21] and the risk of loss of Chinese or even all Asian deliveries due to an incident over Taiwan [22] further raise the value of domestic sources. Even though labor cost in Asian countries is still lower than in the US but in totality, the appeal of US manufacturing is very prominent and obvious [23]. Many survey-based studies were conducted to quantify the benefit and feasibility of reshoring [23-28]. Moreover, researchers also tried to identify the forces that influence reshoring decisions and further implications. This research paper has briefly complied with some of the important research to present the facts, figures, and forces driving the reshoring phenomena. The economic impact [29] on the overall supply chain of reshoring is presented in this research through TCO analysis and recommended for immediate reshoring for those categories of components.

### C. Greenhouse Gas Protocol (GHGP)

Scope 3 GHG emissions will primarily be calculated from activity data such as fuel use or passenger miles and published or third-party emission factors. In most cases, if source- or facility-specific emission factors are available, they are preferable to more generic or general emission factors. Industrial companies may be faced with a wider range of approaches and methodologies. They should seek guidance from the sector-specific guidelines on the GHG Protocol or from their industry associations (e.g., International Aluminum Institute, International Iron and Steel Institute, American Petroleum Institute, WBCSD[8] Sustainable Cement Initiative, and International Petroleum Industry Environmental Conservation Association. The guidance for each calculation tool includes the following sections: (1) The overview provides an overview of the purpose and content of the tool, the calculation method used in the tool, and a process description. (2) Choosing activity data and emission factors provide sector-specific good practice guidance and references for default emission factors. (3) Calculation methods describe different calculation methods depending on the availability of site-specific activity data and emission factors [30]. (4) Quality control provides good practice guidance. (5) Internal reporting and documentation guide internal documentation to support emissions calculations.

### D. Calculation formula - Distance-based method

$CO_2$ emission from transportation =
= sum across transport modes and/or vehicle types

$$= \sum \left\{ \left( \sum_{i=1}^{i=n} M_i \right) \times \left( \sum_{j=1}^{j=m} D_j \right) \right\} \times E_k \qquad (5)$$

Where, M = mass of goods purchased in (tonnes of volume), i = number of different items, D = distance traveled in transport leg in (miles), j = number of different items associated with that travel, E = emission factor of transport mode or vehicle type (kgCO$_2$ e/tonne or volume/mile) coming right from [30]

$CO_2$ Emission from road transport =

$$= \sum \left\{ \left( \sum_{i=1}^{i=n} M_{Road\,(i)} \right) \times \left( \sum_{j=1}^{j=m} D_{Road\,(j)} \right) \right\} \times E_{Road\,(k)} \qquad (6)$$

Where M = mass of goods purchased in (tonnes of volume) that travel on road, $Road_i$ = number of different items on road transport, D = distance traveled in road transport leg in (miles), j = number of different items associated with that road travel, E = emission factor of road transport or vehicle type (kgCO$_2$ e/tonne or volume/mile) coming right from [30]

$CO_2$ Emission from sea transport =

$$= \sum \left\{ \left( \sum_{i=1}^{i=n} M_{Sea\,(i)} \right) \times \left( \sum_{j=1}^{j=m} D_{Sea\,(j)} \right) \right\} \times E_{Sea\,(k)} \qquad (7)$$

Total emissions from transport (upstream) is calculated as
= emission from (road transport + sea transport) (8)

## III. NUMERIC RESULTS

In this section, we will demonstrate the previously explained model in a real-world company. Although the study was done on this company, to maintain confidentiality we will not reveal the name and also use the parts description using more general terms. The ABC company is rooted in innovation and provides world-class anti-vibration and foam products for the automotive industry that improve lives globally. It is importing components/parts from different countries across the world. During the eighties and late nineties, many US manufacturing companies mass-outsourced their operations overseas. The ABC company has been thinking about doing so.

### A. Prioritizing Components for Reshoring Using RI

To understand the company's state, (i.e., where it stands, what are its products, which all parts they are procuring, which parts should be outsourced/reshore, etc.), AS-IS analysis is conducted. It has been observed that ABC company is paying high logistics and tariff costs for their offshore parts. From ABC's projected spending for the fiscal year 2022, we identified that approximately 57% is overseas spending. Approximately 70% of that overseas content is from China, followed by Japan (19%), Thailand (7%), and South Korea (4%). Analysis of the tariff on parts from China highlighted four commodities – Casting products, Stamping products, Forming products, and Mounting products – accounting for most of the tariffs.

### TABLE II
RI %, TRADE DEFICIT, AND LOGISTICS COST % AGAINST THE 6-DIGIT NAICS CODE.

| Items | NAICS Code | RI % | Trade Deficit 100 K | Logistics cost % | Tariff % |
|---|---|---|---|---|---|
| Casting | 331523 | 25 | 55 | 9 | 41.13 |
| Stamping | 336370 | 30 | 14 | 12.89 | 25.31 |
| Forming | 331318 | 23 | 22 | 7.24 | 16.28 |
| Mounting | 331210 | 26 | 0.37 | 9.43 | 13.92 |
| Rubber | 325212 | 22 | -50 | 9.16 | 1.56 |
| Mechanical | 332999 | 20 | 0.009 | 5.25 | 0.90 |
| Plastics | 325211 | 23 | -100 | 10.26 | 0.27 |

In the above table, we figured out the component group's NAICS code, and based on the calculation done [31], we pulled out the data of RI, Trade deficit, and CIF costs. So, understandably, if the RI is high, imports are greater than exports, and there are significant logistics costs involved in that particular industry, that is our suitable candidate for reshoring. If we plot this table in a chart, we will be able to figure out the relation between the three major deciding factors from fig.1.

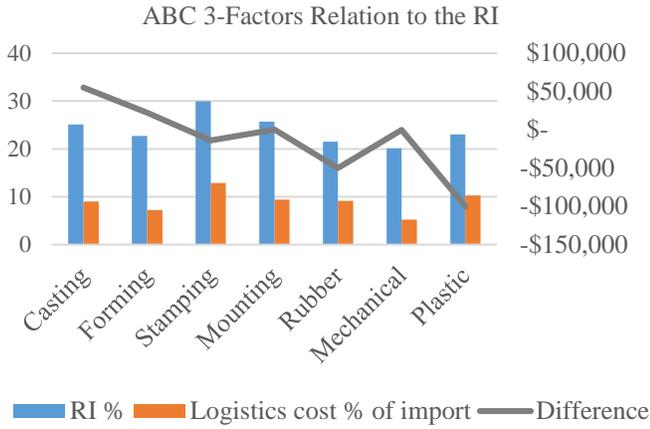

Fig. 1. ABC 3-Factors Relation to the RI

As a part of the evaluation of components for reshoring, reshorability indices concerning trade deficits have been analyzed. The importance of different factors for a particular industry and the strength of a particular country are taken into account to create a 'Reshorability Index'. This Index will provide a comparative benefit if a currently outsourced product is manufactured back in the US from a particular country. The baseline for reshoring is that if the Reshorability index is high, imports are greater than exports, and there are significant logistics costs involved in that particular industry, which is our suitable candidate for reshoring. Also, in AS-IS analysis, those component groups are responsible for 96.64% of the total tariff cost accumulated.

### B. Calculating Total Cost of Ownership (TCO) China vs US

Here in this stage of this study, we would like to calculate the Total Cost of Ownership for our top candidates. This stage requires core data from the industry, and it is specific to the individual company ABC. The idea is, every item has its price in China and US. The lower purchase price (fob) in China is attracting procurement managers, after buying those items there are costs for shipping, packaging, duty fee/custom clearance, and insurance. That cost is summed up in the cost of goods sold bucket. Similarly, in the next bucket, Other hard cost includes; inventory carrying cost, in-transit carrying cost, prototype cost, and so on. Risk and strategic cost are negligible for this study and thus omitted. Finally, depending on the wage increase and price inflation rate in countries, the cost is calculated on a five-year projection. The difference will tell us quantitatively if it is a suitable decision to bring those manufacturing back or not to the USA.

The casting products required in the company ABC incurs the cost below given in table III. As per the TCO analysis, it is apparent that although the purchase price in the USA is higher than China's purchase price by $0.80, the total ownership cost

for the USA product is decreased by $1.65 per item. Also, after five years, the ownership cost difference will be high as $ 2.42 per item. So, it can be inferred that just making a decision solely on the purchase price is misleading.

TABLE III
THE TOTAL COST OF OWNERSHIP (TCO) OF CASTING PRODUCTS, IN CHINA VS THE US

| Cost Factor | The U.S. | China |
|---|---|---|
| FOB price | $ 4.46 | $ 3.66 |
| Total CoGS | $ - | $ 1.67 |
| Total Other Hard Costs | $ 0.04 | $ 0.15 |
| Total Risk Cost | $ - | $ - |
| Total Strategic Cost | $ - | $ - |
| Total Cost Before Freight Premium | $ 4.50 | $ 5.48 |
| 2022 Freight Premium | | $ 0.67 |
| Grand Total Cost of Ownership | $ 4.50 | $ 6.15 |
| Forecast TCO (5 years) | $ 4.70 | $ 7.12 |

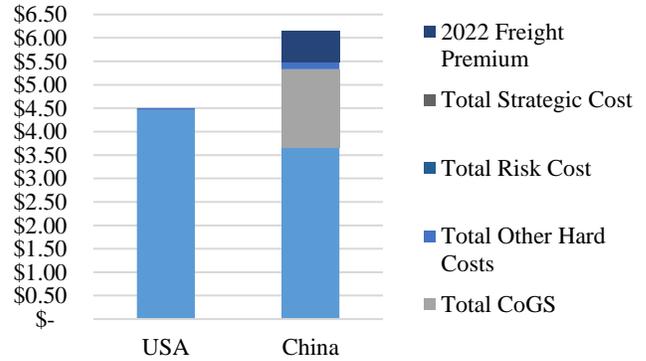

Fig. 2. Casting Products Cumulative Cost by Category (China vs the US)

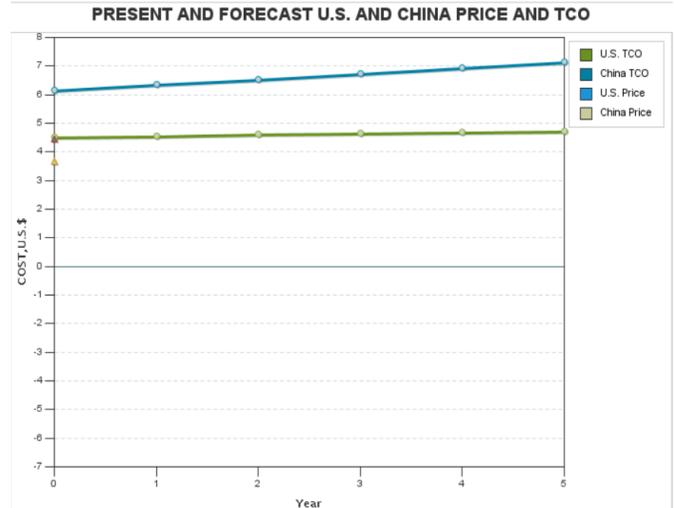

Fig. 3. Casting Products TCO Cost Curve - up to five years, China vs the US

From figures 2 and 3, this is apparent that the TCO is increasing every future year, and at the end of five years, the USA TCO is almost the same as the initial year. But China's TCO is increasing more sharply. So, the company ABC should think about bringing back this manufacturing and start working before losing money.

As far as forming products is concerned, we can see the purchase price difference between the USA and China is only $0.20, and after five years the USA price is even better than the China price, so it is feasible to bring the business back as soon as possible as well.

TABLE IV

TOTAL COST OF OWNERSHIP (TCO) OF FORMING PRODUCTS, IN CHINA VS THE US

| Cost Factor | The U.S. | China |
|---|---|---|
| FOB price | $ 1.12 | $ 0.92 |
| Total Other CoGS | $    - | $ 0.47 |
| Total Other Hard Costs | $ 0.01 | $ 0.03 |
| Total Risk Cost | $    - | $    - |
| Total Strategic Cost | $    - | $    - |
| Total Cost Before Freight Premium | $ 1.13 | $ 1.42 |
| 2022 Freight Premium | | $ 0.67 |
| Grand Total Cost of Ownership | $ 1.13 | $ 2.09 |
| Forecast TCO (5 years) | $ 1.18 | $ 2.43 |

Similarly, if we plot the data from table IV, it would be apparent that the forming products TCO is increasing every future year, and at the end of five years, the USA TCO is better than China because the cost is higher for China. So, the company ABC should choose the suppliers domestically. Not only that it is apparent that from a fob standpoint, China is in the better position, over the other cost factors count, the landed cost has gone competitively less competitive also over the years as well.

The stamping product is a strong candidate for choosing domestic suppliers because of its lower total cost of ownership value when compared to China suppliers. From table V, Although the purchase price is higher, the total cost of ownership is lower for the US. Within five years, the USA suppliers will be significantly cost-saving for the ABC company. The TCO for Stamping products shows that the US suppliers are the best in terms of the purchase cost and during five years the savings will be significantly high based on the future cost trend as well.

TABLE V

TOTAL COST OF OWNERSHIP (TCO) OF STAMPING PRODUCTS, IN CHINA VS THE US

| Cost Factor | The U.S. | China |
|---|---|---|
| FOB price | $ 3.41 | $ 2.79 |
| Total Other CoGS | $    - | $ 1.39 |
| Total Other Hard Costs | $ 0.03 | $ 0.11 |
| Total Risk Cost | $    - | $    - |
| Total Strategic Cost | $    - | $    - |
| Total Cost Before Freight Premium | $ 3.44 | $ 4.29 |
| 2022 Freight Premium | | $ 0.67 |
| Grand Total Cost | $ 3.44 | $ 4.96 |
| Forecast TCO (5 years) | $ 4.00 | $ 5.56 |

From table VI, the Mounting products are also competitive in terms of cost comparison within the US vs China. The purchase cost difference today is $0.17 (China is lower than the USA), but after five years the USA's cost will be $1.14 lower than China's. So, this is a better decision to bring those back to the US soil as a cost-saving, the best parameter for the overseas suppliers is not that competitive as well.

TABLE VI

TOTAL COST OF OWNERSHIP (TCO) OF MOUNTING PRODUCTS, IN CHINA VS THE US

| Cost Factor | The U.S. | China |
|---|---|---|
| FOB price | $ 0.92 | $ 0.75 |
| Total Other CoGS | $    - | $ 0.35 |
| Total Other Hard Costs | $ 0.01 | $ 0.05 |
| Total Risk Cost | $    - | $    - |
| Total Strategic Cost | $    - | $    - |
| Total Cost Before Freight Premium | $ 0.93 | $ 1.15 |
| 2022 Freight Premium | | $ 0.67 |
| Grand Total Cost | $ 0.93 | $ 1.82 |
| Forecast TCO (5 years) | $ 0.97 | $ 2.11 |

The cost comparison below is solely taken from the purchase price and future purchase price in terms of the Total Cost of Ownership (TCO) calculation. Still, domestic production and suppliers are better for bringing jobs back to the USA, and also the company will have a chance to create JIT production systems. Not only that, the "made in the USA" tag is significantly impactful to some industries because recent surveys suggest that 70% of the buyers in the USA are willing to pay 10% more to get the product manufactured in the USA. It is also supported by the Reshorability Index that the Casting and Stamping products are best for reshoring as the TCO suggests because the RI for that industry was higher than others and also, we get a better result in that TCO analysis. So, the TCO analysis is backed up by the Reshoring Index.

TABLE VII

PURCHASE PRICE DIFFERENCE IN CHINA VS THE USA AFTER TCO ANALYSIS

| Product Name | China Advantage on FOB Present day (per unit) | US Advantage on TCO Present Day (per unit) | US Advantage on TCO after 5 years (per unit) |
|---|---|---|---|

| Casting | $0.80 | $1.64 | $2.42 |
| Forming | $0.20 | $0.96 | $1.25 |
| Stamping | $0.62 | $1.53 | $1.56 |
| Mounting | $0.17 | $0.89 | $1.14 |

The above table VII shows the purchase price difference between Chin and US. From the TCO analysis for company ABC, we find out that although the fob price today is lower in china market the added cost for tariff 301, transportation, and duty makes the scenario opposite to what we perceived before. That is why it is crucial to conduct the TCO analysis to find out the actual landed cost of the purchased products. The result also demonstrates that after five years with the inflated price the overseas purchase price will be a record high for the company to be competitive in the market. For example from table VII, the casting products fob today in the Chinese market is 0.80 USD less than that of the USA market for company ABC, but after TCO calculation it is found that the US advantage for the ownership cost of the same product today will be 1.64 USD, more than 50% less from the Chinese products. How did that happen? Well, after considering the tariff and other costs mentioned in the study we have found that the landed cost of the products bought from Chinese markets incurs more cost during the process of getting that in-house in the US. Also, if we take the inflation rate, and wage market rise into consideration, after five years the landed cost will be 125% increased! The other components also show the same trend for the future years.

For longer-term considerations, we broke out Freight Premium which will probably come down substantially over the next few years and Section 301 tariffs may come down eventually, but not soon. TCO favored the U.S. even without the Freight Premium. We did not quantify risk, strategic, and green/ESG costs. Given the tension over Taiwan and Chinese threats to stop shipping key automotive components if the U.S. Innovation and Competition Act is passed, the profitability advantage of shifting to U.S. sources is likely to be understated by the analysis.

### C. Calculating emissions from transportation (Scope 3)

Companies may use the following methods to calculate scope 3 emissions from transportation:

1. **Fuel-based method**: which involves determining the amount of fuel consumed (i.e., scope 1 and scope 2 emissions of transport providers) and applying the appropriate emission factor for that fuel

2. **Distance-based method:** which involves determining the mass, distance, and mode of each shipment, then applying the appropriate mass-distance emission factor for the vehicle used

3. **Spend-based method:** involves determining the amount of money spent on each mode of business travel transport and applying secondary (EEIO)[9] emission factors.

The GHG Protocol has a calculation tool for transportation that uses a combination of fuel-based and distance-based methods. This combination is used because $CO_2$ is better estimated from fuel use, and $CH_4$ and $N_2O$ are better estimated from distance traveled. The tool uses fuel-efficiency ratios to convert either type of activity data (fuel or distance) supplied by the user into either fuel or distance depending on the GHG being calculated. [10]

### D. Distance-based method (transportation)

In this method, distance is multiplied by the mass or volume of goods transported and relevant emission factors that incorporate average fuel consumption, average utilization, average size and mass or volume of the goods and the vehicles, and their associated GHG emissions.

Emission factors for this method are typically represented in grams or kilograms of carbon dioxide equivalent per tonne-kilometer or TEU-kilometer[11]. Tonne-kilometer is a unit of measure representing one tonne of goods transported over 1 kilometer. TEU-kilometer is a unit of measure representing one twenty-foot container equivalent of goods transported over 1 kilometer.

The distance-based method is especially useful for an organization that does not have access to fuel or mileage records from the transport vehicles or has shipments smaller than those that would consume an entire vehicle or vessel.

If sub-contractor fuel data cannot be easily obtained to use the fuel-based method, then the distance-based method should be used. The distance can be tracked using internal management systems or if these are unavailable, online maps. However, accuracy is generally lower than the fuel-based method as assumptions are made about the average fuel consumption, mass or volume of goods, and loading of vehicles.

Companies should collect data on the distance traveled by transportation suppliers. This data may be obtained by:

- Mass or volume of the products sold
- Actual distances provided by transportation supplier (if actual distance is unavailable, companies may use the shortest theoretical distance)
- Online maps or calculators
- Published port-to-port travel distances. The actual distances

---

should be used when available, and each leg of the transportation supply chain should be collected separately. Here in this study, we used the distance-based method [30] to find out the competitive advantage over reshoring. As the transportation data is available with mass-distance traveled, we developed two scenarios where in the first one, the previous supply chain using (road transport-water transport-land transport) is taken into consideration. The distance from China to the port is taken from the average distance; the waterway distance is also averaged out depending on the available distance records. based on the mass required for the ABC company, we took the mass distance. In the last portion, port-to-warehouse distance is also averaged out to find out the total emission for carbone-di-oxide, methane, and nitrous oxide. At the end of the calculation, we find out the total greenhouse gas emission for this supply chain. For the second scenario, we took the average distance from any particular area to the ABC company location and remove the water distance. This scenario also provides the GHG emission which we could now compare with the previous result. It is important to note that we are more interested to find out the impact of reshoring on reducing GHG emissions. As given in below table VIII we have found a clear negative relation, between the reshoring decision and emissions, on the other hand, GHG emissions get reduced by 78% if we change the supply chain by reshoring, which further supports the reshoring initiative positively. We can also see from the data table that 93% of methane gas got reduced and also 88% of dangerous nitrous oxide was reduced from the reshoring decision. Another important point to note here is that we didn't account for the GHG emissions from the port, where all the purchased goods changed the nodes both in China and the US ports. This will further reduce the amount from the value chain. So, from the below study, it is safe to say that, the reshoring decision will improve the supply chain for company ABC in terms of reducing the carbon footprint.

TABLE VIII

GREENHOUSE GAS EMISSIONS REDUCED PERCENTAGE FROM OUTSOURCING TO RESHORING DECISION

| Mode of Transport | Fossil Fuel $CO_2$ (metric tonnes) | $CH_4$ (kilograms) | $N_2O$ (kilograms) |
|---|---|---|---|
| Road | 64% | 64% | 64% |
| Water | 100% | 100% | 100% |
| Total Emissions | 77% | 93% | 88% |
| Total GHG Emission (metric tonnes CO2e) | **78%** | | |

In the below fig. 4, it shows that the reshoring decision reduce the scope 3 GHG emissions from the ABC company's value chain including fossil fuel $CO_2$, $CH_4$, and $N_2O$ respectively by 77%, 93%, and 88%. This summarizes the total 78% of GHG emissions from the entire value chain. From this study and calculations, there are some important factors to be noted: the study was done to find out the GHG emission reductions using the load-distance methods, but there is a significant portion could be saved during the loading and unloading periods both in the US and China ports. The long handling and waiting time in the port at the current time will make the emission significant because those handling equipment are operated using engines.

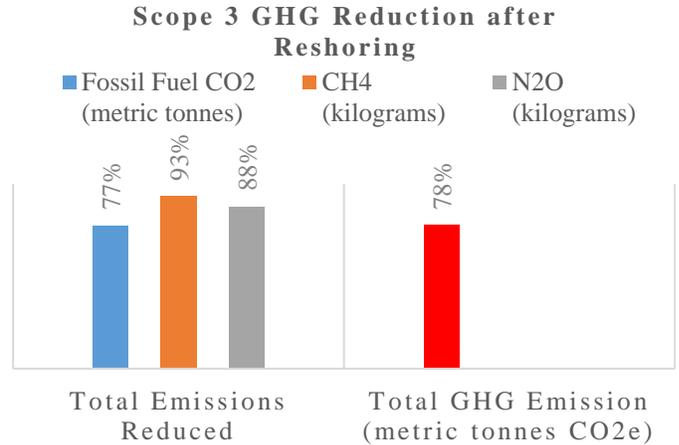

Fig. 4. Scope 3 GHG emission reduction percentage-wise from ABC company's value chain.

## IV. DISCUSSION

The purpose of this section is to analyze and present relevant information on the economic and green impact of reshoring techniques. As we first developed the methods for calculating the reshoring economic impacts, the case study data clearly showed that- different components coming from overseas will be more profitable if purchased domestically. Not only have the profits become multiplied over the years due to the inflation rate and wage increase to the global market, but recent years also show significantly decreased purchased prices in the domestic markets due to the added benefit over the tariff 301 imposed on outsourced products and materials. The other section of the case study shows the positive relation of reshoring decisions with green impacts. In other words, the GHG emission will be reduced from the ABC company's life cycle under scope 3 emission. So, it is apparent that the positive impact of the reshoring decision is twofold: one directly adds value to the company's bottom line in terms of economic profitability and at the same time it also reduces the GHG emission from the company value chain to be able to more competitive with their counterparts.

This case study and the increasing number of reports from companies prove that reshoring is a real trend, and although its magnitude remains in question, its direct and indirect effects can be observed already throughout the United States. Companies have started to reevaluate the profitability of operating overseas and are considering relocating back to North America. Supply chain disruptions and instability have hastened the reshoring trend. The pandemic exposed every global supply chain crack. Not only has estimating shipping time become unpredictable, but costs also skyrocketed when the demand came rushing back.

Reduced capacity caused by labor shortages and elevated oil prices has also made transpacific shipping difficult. The COVID-19 pandemic also spurred new interest in reshoring as global supply chains became disrupted by outbreaks and lockdowns. In fact, for 2023 reshoring activities were expected to hit record highs.

This case study was done on company ABC which purchased raw materials from overseas and those are not heavy/bulk materials, they are relatively at the lower end of the weights generally transported from overseas. Both from the reshoring index and TCO analysis, the weight of the transport materials could make a significant difference in the result. Also in the second part of the analysis, we did our GHG emission calculations based on the weight-distance method where the weight plays an important role in deciding the emission factors. It is always meaningful and important to get the recent TCO done for every industry which solely depends on outsourcing. This study would be a guideline to conduct those studies to make an important decision for any particular manufacturing industry. So it is recommended that if a company depends on outsourcing, this is high time to conduct the TCO analysis also GHG emission analysis to get a complete picture of the entire scenario of savings and positive impacts. If a medium-sized company like ABC ended up saving millions of dollars and 78% of emissions just incorporating the reshoring decision, the large companies will see considerable differences and savings from their value chain in both ways.

## V. CONCLUDING COMMENTS AND DIRECTIONS FOR FUTURE RESEARCH

In this paper, we presented a series of models that illustrate how TCO and carbon footprint considerations could be incorporated into business operations. We showed how various outsourcing factors can be incorporated into the reshoring decision-making process. This paper also highlights the impact of operational (reshoring) decisions on TCO and carbon emissions and the extent to which adjustments to operations can mitigate emissions. The results show that operational adjustments alone can lead in some cases to significant cost savings and greenhouse gas emission reductions. The results also show the full impact of reshoring in terms of what products or components to reshore or not and how that impacts companies' bottom line. More significantly, the results highlight the opportunity that exists to leverage collaboration across the supply chain to mitigate the emissions; they also highlight important implications of such collaboration (e.g., bringing back the manufacturing jobs to the US soil, in minimized costs and emissions reductions). Operational decisions such as reshoring can improve both economic and environmental sustainability for companies.

Avenues for future research are numerous. As highlighted in this paper, there are numerous facets to how environmental concerns and bringing business back might affect the value chain and how profitable that would be within the business bottom line. The analysis we carried out in this paper highlights some of these interactions. However, each of the issues raised in the paper, as well as others, is worthy of more comprehensive and more rigorous treatment. For instance, a company would be interested to know the exact impact of reshoring on labor costs, supply chain costs, intellectual property costs, etc. Our research was primarily limited to the aggregate impact on TCO.

The scope of this research was limited to the rubber manufacturing industry only. A similar analysis could be carried out using the same model in other industries such as the transportation equipment industry, furniture industry, and electronics industry. The scopes are particularly discussed in the reshorability index (RI) based on specific NAICS codes. Using these industries, it might be possible to characterize analytically the impact of reshoring on GHG emissions and that will also be interesting to see how that complements the findings.

In this paper, we have focused on decisions regarding reshoring and GHG emissions. There are of course other supply chain decisions that are affected by concerns for carbon emissions, including nearshoring (e.g., Mexico, Canada), NAFTA zones, and TPP agreement countries, among many others.

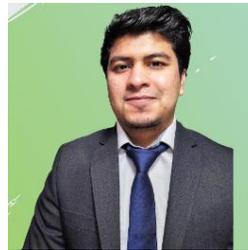

**MD Parvez Shaikh** received a B.Sc. degree in Industrial and Production Engineering from Shahjalal University of Science and Technology, Sylhet, Bangladesh, in 2018. He is pursuing his M.Sc. degree in Logistics Systems Engineering at Bowling Green State University, Bowling Green, Ohio aspires to be graduated in 2023.

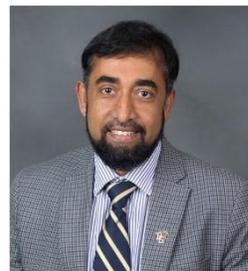

**Dr. Sarder** is a chair and professor of systems engineering at Bowling Green State University (BGSU). Before joining BGSU, he worked in the U.S. Air Force Academy as a distinguished research fellow and as an associate professor and program director at the University of Southern Mississippi. He earned his Ph.D. in industrial and systems engineering from the University of Texas at Arlington. Dr. Sarder authored five books, seven book chapters, and over 100 scholarly articles predominantly in logistics and supply chain. He was awarded more than $3.5M in funding from various agencies including the National Science Foundation and the U.S. Department of Transportation. He served as a PI for multiple reshoring projects and published several articles on the same topic.